# Magnetic and electronic Co states in layered cobaltate $GdBaCo_2O_{5.5-x}$

M. García-Fernández[1], V. Scagnoli[1, 2], U. Staub[1], A. M. Mulders[1,3, 4], M. Janousch[1], Y. Bodenthin[1], D. Meister[1], B. D. Patterson[1], A. Mirone[2], Y. Tanaka[5], T. Nakamura[6], S. Grenier[7], Y. Huang[8], K. Conder[9]

[1]Swiss Light Source, Paul Scherrer Institut, CH-5232 Villigen PSI, Switzerland

[2]European Synchrotron Radiation Facility, BP 220, 38043 Grenoble Cedex 9, France

[3]Department of Imaging and Applied Physics, Curtin University of Technology, Perth, WA 6845, Australia

[4]The Bragg Institute, Australian Nuclear Science and Technology Organization, Lucas Heights, NSW 2234, Australia

[5]RIKEN, SPring-8 Center, Harima Institute, Sayo, Hyogo 679-5148, Japan

[6]SPring-8/ JASRI, Mikazuti, Sayo, Hyogo 679-5198, Japan

[7]Institut Néel, CNRS & Université Joseph Fourier, BP 166, F-38042 Grenoble Cedex 9, France

[8]Van der Waals–Zeeman Instituut, Universiteit van Amsterdam, Valckenierstraat 65, 1018 XE Amsterdam, The Netherlands

[9]Laboratory for Neutron Scattering, Paul Scherrer Institut & ETH Zürich, CH-5232 Villigen PSI, Switzerland

## Abstract

We have performed non-resonant x-ray diffraction, resonant soft and hard x-ray magnetic diffraction, soft x-ray absorption and x-ray magnetic circular dichroism measurements to clarify the electronic and magnetic states of the $Co^{3+}$ ions in $GdBaCo_2O_{5.5}$. Our data are consistent with a $Co^{3+}_{Py}$ HS state at the pyramidal sites and a $Co^{3+}_{Oc}$ LS state at the octahedral sites. The structural distortion, with a doubling of the *a* axis ($2a_p$ x $2a_p$ x $2a_p$ cell), shows alternating elongations and contractions of the pyramids and indicates that the metal-insulator transition is associated with orbital order in the $t_{2g}$ orbitals of the $Co^{3+}_{Py}$ HS state. This distortion corresponds to an alternating ordering of *xz* and *yz* orbitals along the *a* and *c* axes for the $Co^{3+}_{Py}$. The orbital ordering and pyramidal distortion lead to deformation of the octahedra, but the $Co^{3+}_{Oc}$ LS state does not allow an orbital order to occur for the $Co^{3+}_{Oc}$ ions. The soft x-ray magnetic diffraction results indicate that the magnetic moments are aligned in the *ab* plane but are not parallel to the crystallographic *a* or *b* axes. The orbital order and the doubling of the magnetic unit cell along the *c* axis support a non-collinear magnetic structure. The x-ray magnetic circular dichroism data indicate that there is a large orbital magnetic contribution to the total ordered Co moment.

## I. INTRODUCTION

Perovskites ($ABO_3$) and related materials are of fundamental importance, due to their rich electronic, magnetic and structural properties. An interesting question concerns the origins and driving forces of the metal-insulator (MI) transitions in these materials. Systems where a MI transition occurs as a function of temperature are manganites, titanates and nickelates [1]. Charge, magnetic and orbital ordering phenomena are known to play a crucial role in the carrier localization that occurs at these transitions. Still the individual significance of these order parameters at a microscopic level is not well understood. Accurate knowledge of the crystal structure, in particular in the electronically localized (insulating) phase, is important for the determination of the electronic states of the transition metal ions. The building blocks of transition metal perovskites are oxygen octahedra, with the transition metal ion at the center separated by interstitial cations forming a cubic lattice. The deformation of the oxygen octahedra from cubic symmetry and alignment, together with site symmetry of the transition metal ion, is used to determine the orbital ground state, and it often gives valuable information on possible magnetic structures.

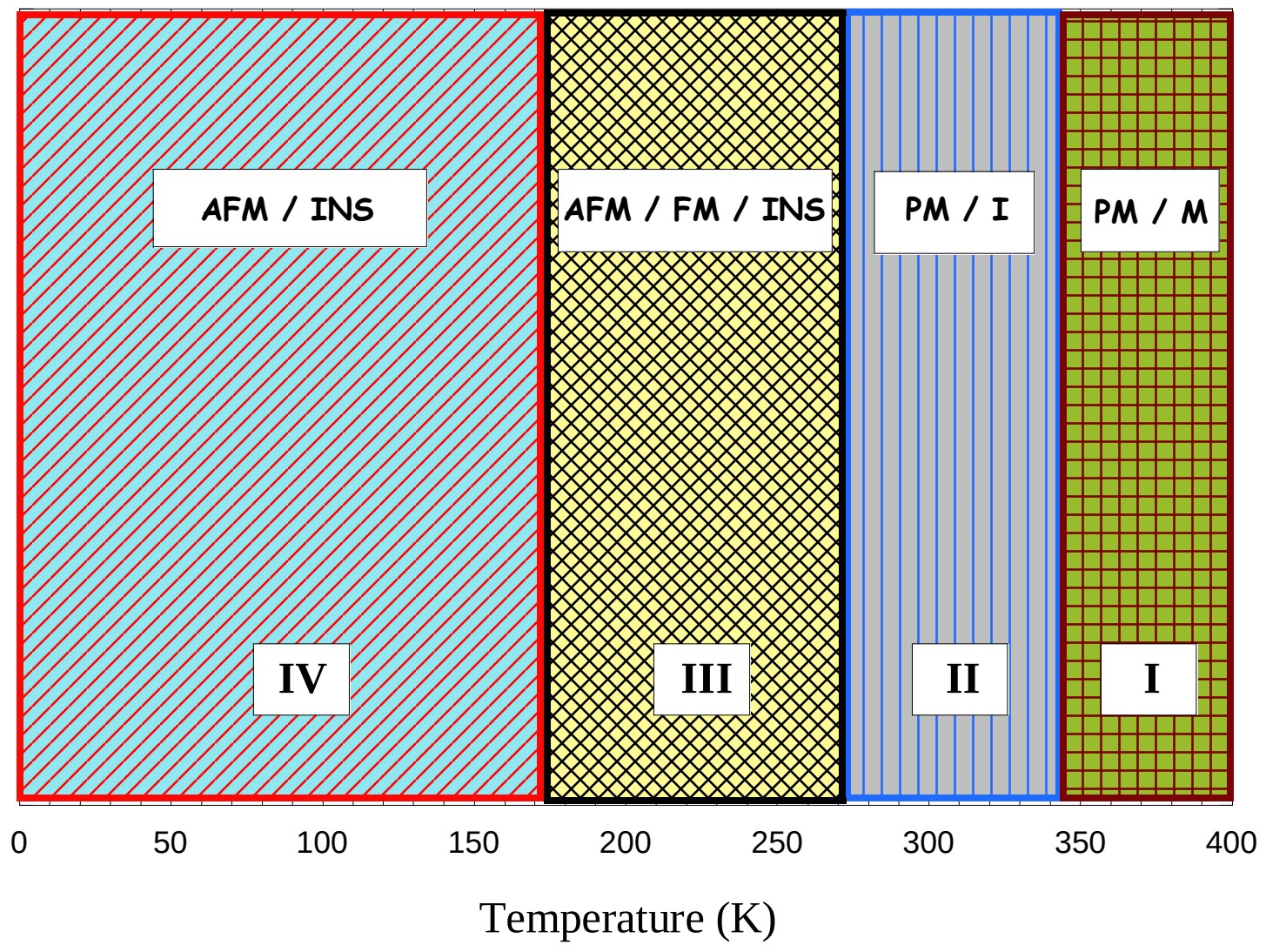

**Figure 1.** Schematic phase diagram of the $RBaCo_2O_{5.5}$ compounds. Four different phases are shown as a function of temperature. AFM stands for antiferromagnetic, FM for ferromagnetic and PM for paramagnetic, while I is insulator, and M metal.

For the case of Co based perovskites, in addition to the magnetic and orbital degrees of freedom, $Co^{3+}$ ions can either be in a low-spin (LS, S=0), an intermediate-spin (IS, S=1) or a high-spin state (HS, S=2). The spin state of the Co ions depends on the symmetry and strength of the crystal field and the intra-atomic exchange energies. The prototypical material having different spin-states is $LaCoO_3$. The determination of the spin state of the Co ions in the different electronic phases of $LaCoO_3$ is challenging and still of current interest [2-4]. The $R\mathrm{BaCo_2O_{5.5-x}}$ (*R*=rare earth) compounds are related perovskite materials for which spin states and possible transitions and ordering schemes have been proposed [5-12]. Here we are concerned about the material with $x \approx 0$, for which a generic phase diagram (minimal amount of possible phases) can be drawn (see Figure 1). The high temperature phase is metallic and paramagnetic and labeled with I. Around 350 K (for *R*=Gd) a metal insulator transition occurs making the material a paramagnetic insulator at room temperature.

Associated with the MI transition, structural distortions and changes in the magnetization are observed [5, 6]. Susceptibility ($\chi$) measurements show a significant increase of slope in $\chi$/T when cooling through the MI transition from which a reduced magnetic moment in the Curie-Weiss fit was extracted. This reduction in moment was interpreted as a $e_g$ change from HS for $T>T_{MI}$ to IS for $T<T_{MI}$ [13].

By further cooling below room temperature, a magnetic transition sets in observed by a strong increase of the magnetization indicative for a significant ferromagnetic component (phase III). As neutron diffraction observed for R=Tb, Nd [6, 7] simultaneously antiferromagnetic reflections, this phase may therefore be ferrimagnetic or of a canted antiferromagnetic type. Around 200 K (strongly depending on the exact oxygen content), this ferromagnetic component starts disappearing and the material enters an antiferromagnetic insulating phase (phase IV).

These materials also exhibit the largest magneto-resistance (MR) found so far in Co based oxides. The MR is linked to magnetic phase transitions and to the competition between antiferromagnetic and ferromagnetic components in phase III / IV [15], that occur below room temperature.

The crystal structure for $x \approx 0$ is based on a layering of the Ba and Gd ions, separated by $CoO_2$ planes. The Ba planes are fully oxidized, whereas for the Gd layer, chains of oxygen along the crystallographic *a* axis are formed, leading to an approximately ($a_p \times 2a_p \times 2a_p$) ($a_p$ cubic perovskite lattice constant) unit cell. Note that the fully oxidized (x= -1/2) material corresponds to the simple A-site ordered perovskite $ABO_3$ structure; with the A site layering of Ba and Gd. For $x \approx 0$, the high temperature crystal structure, exhibits Pmmm symmetry with two distinct Co sites [14]. The Co ions on both sites are predicted to be trivalent. One of the Co is located inside a square pyramid (five-fold coordinated), and the other is inside a slightly distorted octahedron (six-fold coordinated), as shown in Fig. 2. Both Co ions have relatively low symmetry, with a single 2 fold axis. For the low temperature, insulating state, various models have been proposed for the magnetic, spin, and orbital ordered ground state [5-12]. Most of the studies assume that the square pyramidal Co is IS, and that the octahedral Co is either fully or partly LS or that there is at least an admixture of LS in the ground state (see e.g. [6, 7]). The determination of the electronic and magnetic states, in particular that of the Co spin state, is hindered by several factors. The oxygen non-stoichiometry has been shown to vary depending on the particular rare earth ion as well as preparation conditions, which leads to different ratio of admixtures of both $Co^{2+}$ and $Co^{4+}$ ions compared to the main $Co^{3+}$ states. This significantly affects the electronic properties of the $R\mathrm{BaCo_2O_{5.5-x}}$ materials. The electronic ground state can often be inferred from details of the crystal structure, as well as from the magnetic structure. Upon synthesis in air, the compound $GdBaCo_2O_{5.5-x}$ has an x-value close to 0.5. Gd is a strong neutron absorber, and this limits an accurate determination of the magnetic structure with neutron scattering. In addition, the growth of high-quality crystals is very difficult, and the extraction of microscopic information from polycrystalline materials is limited, in particular due to the presence of two differently coordinated Co ions.

In this paper, we present a synchrotron based single crystal study using high-energy x-ray diffraction, resonant magnetic soft and hard x-ray diffraction and soft x-ray magnetic absorption and circular dichroism (XMCD). The high-energy diffraction data show the development of additional structural reflections at T< $T_{MI}$ that are indexed with (*h*/2 *k* *l*). The minimal low temperature structure is therefore ($2a_p$ *x* $2a_p$ *x* $2a_p$). The resonant magnetic soft x-ray diffraction experiments observed the (0 0 ½) reflection, which is consistent with a doubling of the magnetic unit cell along the c axis in the antiferromagnetic (AFM) phase IV. These two results suggest that either orbital, charge (disproportionation) or spin-state ordering occurs at $T_{MI}$ within the *ac* plane. The azimuthal angle dependence is not well described by a simple magnetic AFM structure with magnetic moments pointing along a certain crystallographic axis. The XMCD data indicate that the Co ions have a large orbital magnetic moment. Consequences of our observations are discussed in view of the various proposed models describing the electronic state of $GdBaCo_2O_{5.5-x}$.

## II. EXPERIMENTS

A single crystal of $GdBaCo_2O_{5.5-x}$ has been grown by the traveling floating-zone method in the mirror furnace of the University of Amsterdam. The crystal has been characterized with Cu *K*α radiation: the crystal is *a, b* twinned with approximately equal populations of the two twins, when integrated over the intrinsic volume of the crystal. The susceptibility measurements show two magnetic transitions at T=270 K (from phase II to III) and 180 K (from phase III to IV), confirming that the non-stoichiometry equals approximately x=0.05, according to Ref. [15]. High-energy x-ray diffraction experiments have been performed at the BL19LXU beamline at SPring-8 in Japan. An incident energy of 30 keV was selected by a double crystal monochromator, and a Pt coated mirror was used for harmonic suppression. The beam size was 0.2 x 0.2 mm. Resonant x-ray diffraction experiments at the Co *K* edge (~7.7 keV) were performed using the 2+3 circle diffractometer of the Material Sciences beamline at the Swiss Light Source of the Paul Scherrer Institut in Switzerland.

The (0 0 6) reflection of graphite was used for polarization analysis and, a solid state pixel detector (Pilatus I Ref. [16]) based on Si technology was used to obtain integrated intensities without polarization analysis. With a closed cycle refrigerator, temperatures between 30 and 400K were investigated. Resonant magnetic soft x-ray diffraction experiments have been performed at the RESOXS endstation at the SIM beamline of the Swiss Light Source at the Paul Scherrer Institut. Measurements were performed in horizontal scattering geometry at the Co $L_{2,3}$ edges, between 30 and 300 K, using a helium-flow cryostat. The linear polarization of the incident radiation was either horizontal ($\pi$) or vertical ($\sigma$). Polarization analysis of the scattered radiation was performed using a graded W/C multilayer [17]. Horizontal and vertical mounting of the analyzer stage allowed for $\pi$' and $\sigma$' detection of the scattered radiation. The XMCD experiment was performed at the twin helical undulator beamline BL25SU of SPring-8. The measurements were performed in magnetic fields from 0 to 1.9 Tesla produced by an electro-magnet. The XMCD data was recorded in total electron yield mode using 1 Hz helicity switching. The degree of circular polarization was 96%. A single crystal of $GdBaCo_2O_{5.5-x}$ was cooled down to 200K (phase III) using a continuous helium flow cryostat. The field direction pointed 45 degrees away from the *c*-axis. The beam was incident on the sample surface, making an angle of 10 degrees with the applied magnetic field.

## III. CRYSTAL STRUCTURE DETERMINATION

Recently, the low temperature crystal structure for $R$BaCo$_2$O$_{5.5-x}$ (x≈0), has been proposed to undergo a doubling along the crystallographic *a* axis as indicated by the observation of (*h*/2 *k l*) structural superlattice reflections in $TbBaCo_2O_{5.5-x}$. The crystal was proposed to be in the ($2a_p$ x $2a_p$ x $2a_p$) cell or even ($2a_p$ x $2a_p$ x $4a_p$), inferred from symmetry considerations of the magnetic structure [7]. The proposed symmetry is Pmma and 11 superlattice reflections have been used to refine some of the atomic positions for the *R*=Gd case [10]. The number of observed reflections is not sufficient to determine all the atomic displacements with respect to the high temperature Pmmm crystal structure. For the Tb analogue, high-resolution neutron powder diffraction data were analyzed in the ($2a_p$ x $2a_p$ x $2a_p$) (Pmma) structure; although no superlattice reflections were observed.

We have collected x-ray diffraction data on different (*h k l*) reflections for *h* odd at T=10K. (From here on the ($2a_p$ x $2a_p$ x $2a_p$) cell is used.) Due to strong absorption, we used reflection geometry which, combined with the restricted tilt motion of the sample, limited the number of recorded reflections. We recorded in total 34 superlattice reflections of 5 different types: (*h*01), (*h*03), (*h*02), (*h*11) and (-1*k*1). The diffracted intensities were corrected for scattering volume, Lorentz factor and the polarization; the results are shown in Figure 3. The reflections (*h*01), (*h*02), (*h*03) and (*h*11) are obtained from one domain and the (-1*k*1) from the twin domain. The experimental geometry, with the *a* (*b*) axis perpendicular to the crystal surface, and the high momentum resolution of the synchrotron experiment allow the reflections originating from the two twin domains to be distinguished, in contrast to the situation in Ref. [10]. The intensities of symmetry-related reflections such as (400) and (040) were equal, indicating a 50-50 population of the two twin domains within the probed volume (the probing depth was of the order of 40 μm ($\lambda \approx$ 0.4133 Å) ).

A previous diffraction study on $GdBaCo_2O_{5.5-x}$ [10] determined the basic atomic positions and 8 independent displacements via the intensities of 11 superstructure reflections. Due to this rather low number of measured reflections, assumptions about the displacements needed to be made, and this introduced uncertainties; the authors stated that the displacements may be underestimated. We attempted to treat our data in terms of the proposed displacements, but the results were unsatisfactory. Later on, a more accurate measurement on $TbBaCo_2O_{5.5-x}$ was published [7], which proposed an alternate set of displacements. It is generally assumed that the structural distortion for the $R$BaCo$_2$O$_{5.5}$ compounds varies smoothly with $R$. Therefore we have analyzed our superlattice reflections with scaled parameters obtained from the Tb analogue. These scaled displacements also do not describe our data satisfactorily, and we have used them as a starting point to fit the actual atomic displacements. We have performed this analysis on the basis of the unit cell ($2a_p \, x \, 2a_p \, x \, 2a_p$), with the symmetry Pmma (Z=4).

The structure factor of the 5 different types of reflections is developed into a series expansion up to second order in the displacements ($\zeta$, $\phi$, $\delta$) from the mean atomic positions **d**=(x$_0$, y$_0$, z$_0$), (see table 1 for labeling criteria):

$$F(h0l_{odd}) = 4(\pi l)^2 \cdot f_{Co}(Q)\cdot(-1)^{\frac{h-1}{2}}\cdot(-1)^{\frac{l-1}{2}}\cdot \mathrm{A} - 4\pi\cdot f_O(Q)\begin{bmatrix}\left[2l\cdot(-1)^{\frac{h-1}{2}}\cdot\left[\pi l\cdot \mathrm{D}\cdot\sin\left(2\pi l\cdot z_0^{O6}\right)+\mathrm{E}\cdot\cos\left(2\pi l\cdot z_0^{O6}\right)\right]\right]\\ +\left[l\cdot(-1)^{\frac{h-1}{2}}\cdot \mathrm{C}^- - 2h\cdot(\mathrm{F}+\mathrm{G})\right]\end{bmatrix} \quad (1)$$

$$F(h0l_{even}) = 8\pi\cdot f_{Co}(Q)\cdot(-1)^{\frac{h-1}{2}}\cdot\mathbf{B} - 4\pi\cdot f_O(Q)\cdot\begin{bmatrix}2l\cdot(-1)^{\frac{h-1}{2}}\cdot\left[\pi l\cdot\mathbf{D}\cdot\sin\left(2\pi l\cdot z_0^{O6}\right)+\mathbf{E}\cdot\cos\left(2\pi l\cdot z_0^{O6}\right)\right]\\ -\left[l\cdot(-1)^{\frac{h-1}{2}}\cdot\mathbf{C}^+ + 2h\cdot(\mathbf{F}-\mathbf{G})\right]\end{bmatrix} \quad (2)$$

$$\begin{aligned}F(h1l) &= 4\pi^2\cdot f_{Co}(Q)\cdot(-1)^{\frac{h-1}{2}}\cdot \mathrm{A}+\\ &+4\cdot f_O(Q)\cdot\left[\left[\pi\cdot\left[(-1)^{\frac{h-1}{2}}\cdot\mathrm{C}^- + 2h\cdot(\mathrm{G}-\mathrm{F})\right]\right] - (-1)^{\frac{h-1}{2}}\cdot\begin{bmatrix}\cos\left(2\pi\left(y_0^{O6}+\phi_{O6_1}\right)\right)\cdot\sin\left(2\pi l\left(z_0^{O6}+\delta_{O6_1}\right)\right)+\\ \cos\left(2\pi\left(y_0^{O6}+\phi_{O6_2}\right)\right)\cdot\sin\left(2\pi l\left(z_0^{O6}+\delta_{O6_2}\right)\right)\end{bmatrix}\right]\end{aligned} \quad (3)$$

$$F(\bar{1}\;k\;1) \propto -4\cdot f_O(Q)\cdot(-1)^{\frac{h-1}{2}}\cdot\left[\left(2\pi k\delta_{O6_2}\right)-\left(2\pi k\delta_{O6_1}\right)\right]\cdot\sin\left(2\pi k y_0^{O6}\right)\cdot\sin\left(2\pi l z_0^{O6}\right)+\mathrm{M}(O,Co) \quad (4)$$

where $\mathbf{A} = (\delta^2_{CoPy_1} - \delta^2_{CoPy_2} + \delta^2_{CoOc_1} - \delta^2_{CoOc_2})$; $\mathbf{B} = \left(\delta_{CoPy_1} + \delta_{CoPy_2} + \delta_{CoOc_1} + \delta_{CoOc_2}\right)$; $\mathbf{C}^- = (\delta_{O3} - \delta_{O2})$; $\mathbf{C}^+ = (\delta_{O2} + \delta_{O3})$; $\mathbf{D} = \left(\delta^2_{O6_2} - \delta^2_{O6_1}\right)$; $\mathbf{E} = \left(\delta_{O6_1} + \delta_{O6_2}\right)$; $\mathbf{F} = \left(\zeta_{O4}\cdot\sin\left(2\pi l\cdot z_0^{O4}\right)\right)$; and $\mathbf{G} = \left(\zeta_{O5}\cdot\sin\left(2\pi l\cdot z_0^{O5}\right)\right)$

$f_{Co}$ and $f_O$ are the atomic form factors for cobalt and oxygen, respectively, and M(Co,O) represents the contribution from the remaining cobalt and oxygen ions contributing to the $(\bar{1}k1)$ type reflection. We use the same atomic form factors for all Co sites, though in reality they will slightly differ (e.g. different spin-states and different admixtures of $Co^{2+}$ due to oxygen non-stochiometry). We find that M(Co,O) is negligible, as a sinusoidal function describes the experimental data (Figure 3) reasonably well.

Also in this study, the number and distribution of superlattice reflections is insufficient to determine a unique solution, but there are certain requirements that the 12 atomic displacements have to simultaneously fulfill, in order to describe our data. These requirements are sufficient to discard certain possibilities, and we present here a set of displacements that satisfactorily describes our data.

The *l*-dependence of the reflections of the type $(h0l_{odd})$ show an "up/down/up/down" modulation in intensity, (see Figure 3), while the *l*-dependence of the $(h0l_{even})$ reflections do not show such modulation. This implies that the displacements of the oxygen ions O4 and O5 (terms labeled **F** and **G)** have to be of the same sign, i.e. in the same direction (as in Ref. 7, but smaller). The O4 and O5 displacements proposed in ref [6] are the same order of magnitude as ours, but they have opposite sign.

The "up/down/up/down" modulation is represented mathematically by $(-1)^{\frac{h-1}{2}}$ and is associated with the displacement of cobalt and the oxygen ions O2, O3 and O6. For a reasonable description of our data, it is required that the modulation introduced by these terms vanishes for $(h0l_{even})$. This can be accomplished in two different ways.

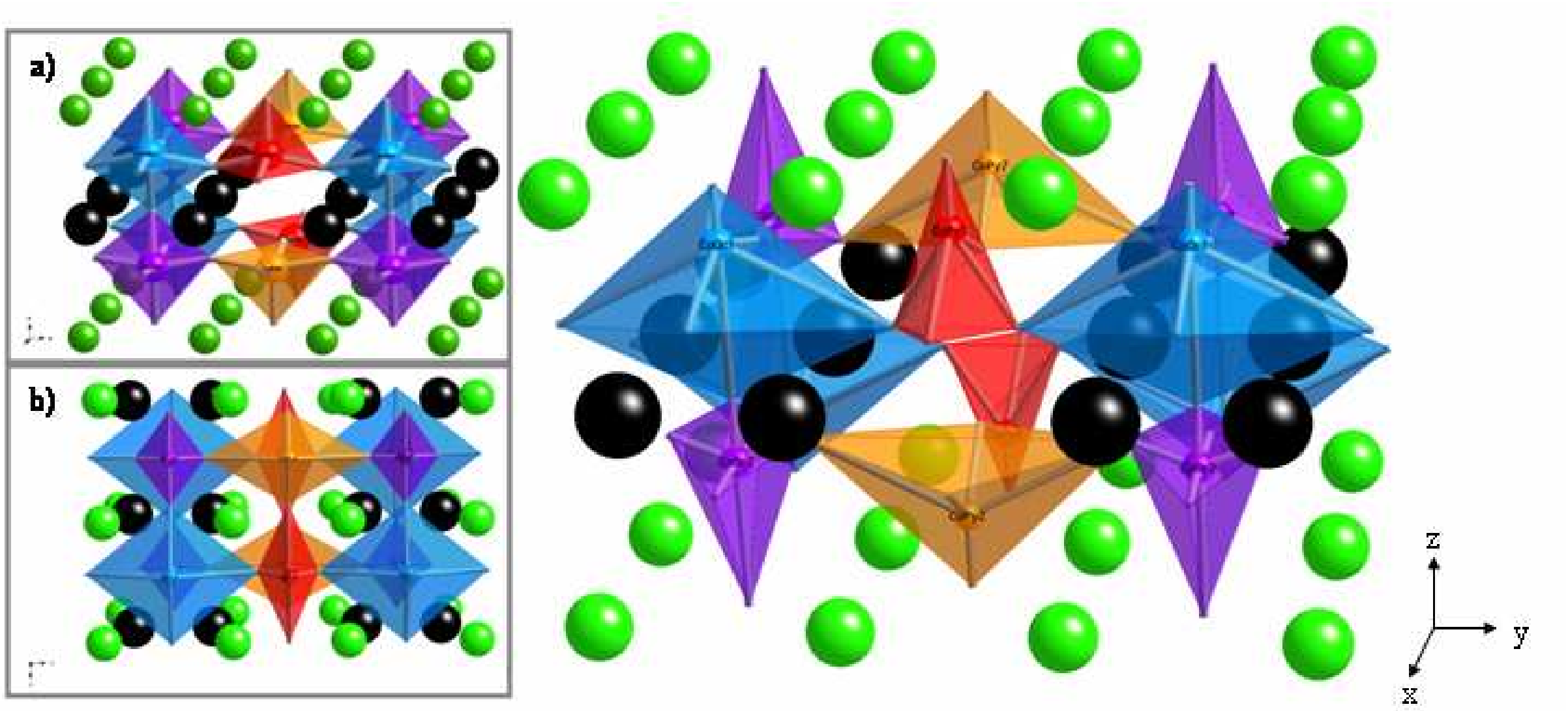

**Figure 2**. Illustration showing the distortion from the Pmmm unit cell in the low temperature phase of $GdBaCo_2O_{5+\delta}$. The displacements have been magnified by a factor of 10 for the sake of clarity. The inset b) shows the top view from the distorted unit cell while inset a) shows the undisturbed "Pmmm" crystallographic unit cell. The colour correspondences are as follows: Ba ions in green, Gd in black and for the cobalt ions we distinguish between sites: $Co_{Oc1}$ in blue, $Co_{Oc2}$ in violet, $Co_{Py1}$ in red and $Co_{Py2}$ in yellow.

The atomic displacements may be so small that each of the individual contributions to the $(-1)^{\frac{h-1}{2}}$ term of the structure factor becomes negligible. However, the O6 contribution plays an important role in the description of the ($h0l_{odd}$) and ($h1l$) reflections and is even the dominant term for describing the (-1$k$1) reflections, so it cannot be zero.

Alternatively, the combination of the atomic displacements may be such that the $(-1)^{\frac{h-1}{2}}$ term is negligible. The ($h0l_{even}$) structure factor depends linearly on the Co atomic displacements (term **B**), whereas for the structure factor of the ($h0l_{odd}$) and ($h$11) reflections, this dependence is quadratic (term **A**). Therefore, the magnitudes of these displacements may be such that the sum of terms related to Co, O2, O3 and O6 becomes insignificant for ($h0l_{even}$), while it remains important for the other type of reflections.

Additionally, we have used FullProf [18] to refine the data, but this did not result in a satisfactory solution, due to the limited range of reflections. Therefore the program was essentially used to crosscheck our results.

| ATOM | POSITION | $x_0/A_1+\zeta$ | $y_0/A_2+\Phi$ | $z_0/A_3+\delta$ |
|---|---|---|---|---|
| Gd | 4h | 0 | 0.2678 + 0.0058 | 0.5 |
| Ba | 4g | 0 | 0.2480 – 0.0012 | 0 |
| $Co_{Py1}$ | 2e | 0.25 | 0 | 0.253 – 0.009 |
| $Co_{Py2}$ | 2e | 0.25 | 0 | 0.747 + 0.008 |
| $Co_{Oc1}$ | 2f | 0.25 | 0.5 | 0.255 – 0.006 |
| $Co_{Oc2}$ | 2f | 0.25 | 0.5 | 0.745 – 0.004 |
| O1 | 2e | 0.25 | 0 | 0.000 |
| O1' | 2e | 0.25 | 0 | 0.5 |
| O2 | 2f | 0.25 | 0.5 | 0.000 + 0.006 |
| O3 | 2f | 0.25 | 0.5 | 0.500 + 0.006 |
| O4 | 4i | 0 - 0.005 | 0 | 0.3092 |
| O5 | 4j | 0 - 0.005 | 0.5 | 0.2694 + 0.002 |
| O61 | 4k | 0.25 | 0.2414 – 0.011 | 0.2960 + 0.009 |
| O62 | 4k | 0.25 | 0.2414 + 0.0109 | 0.7040 + 0.006 |

**Table 1.** Atomic positions (x, y, z) and displacements (ζ, Φ, δ) in fractional coordinates of the Pmma low temperature structure ($2a_p x 2a_p x 2a_p$), obtained from a fit to the intensities of the superlattice reflections recorded with 30 keV x-rays.

The values obtained for the atomic displacements are shown in Table 1, and in Figure 2 we illustrate the corresponding deviation from the *Pmmm* high temperature structure, with an exaggeration of the displacements by an order of magnitude, for clarity.

The distortions are of two types: First, the square pyramids show an alternating contraction and elongation along the *a* and *b* axes. This can be attributed to the Jahn-Teller distortion caused by an orbital order, which can occur with an IS state, based on one electron in the $e_g$ orbital, or a HS state, based on the 4 electrons in the $t_{2g}$ orbitals. The distortion of the octahedra similarly alternates along *a* and *c*, and in addition, the $Co_{Oc1}$ octahedron is enlarged compared to that of the $Co_{Oc2}$.

Lacking further information, such distortions can be attributed to either orbital or to spin state order; the ionic radius of $Co^{3+}$ is larger in the HS state than in the LS state.

Finally, we would like to mention that we did not observe any reflection of type (*h k* l/2) with *h* odd as recently proposed for the Bmmm crystal symmetry indicative for the ($2a_p$ x $2a_p$ x $4a_p$) unit cell in phase III / IV [19]. In case this would be the correct crystal symmetry, the distortion from Pmma must be significantly smaller for our sample than reported in [19].

In summary, the observed structural distortion in the *ac* plane is indicative of an orbital order of $Co^{3+}$ states in the square pyramid, which is discussed when more information on the spin state is available, in section VII.

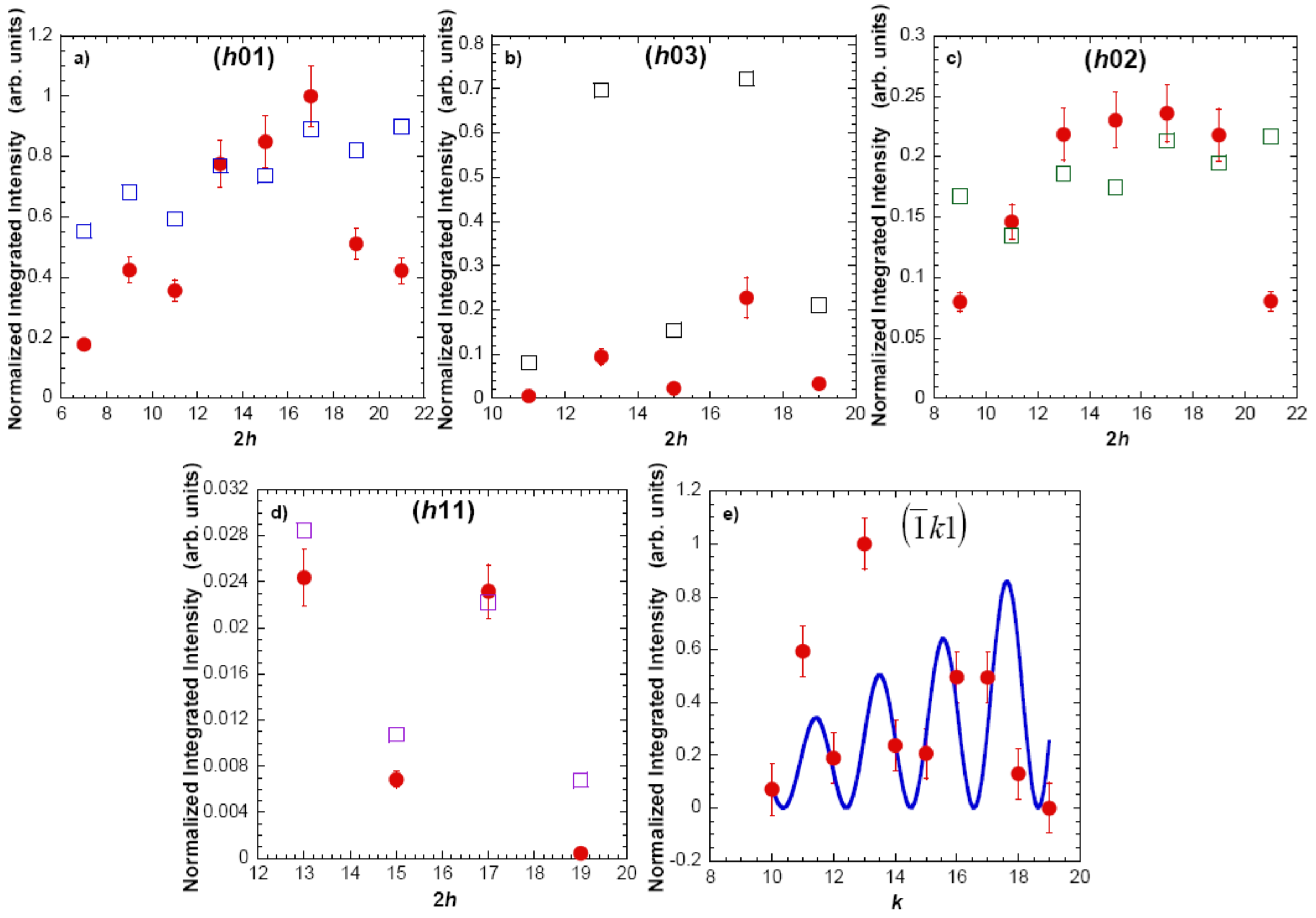

**Figure 3.** Integrated intensities of superlattice reflections recorded with 30 keV x-rays (filled circles), compared to the calculated values from displacements listed in Table 1 (open squares or solid curve).

## IV. CHARGE ANISOTROPIES AROUND Co ATOMS.

Diffraction experiments with x-rays of energies near the Co *K* edge (7.7 keV) are sensitive to the electronic ordering of the Co ions. The Co 4*p* states are directly probed by the dipole transition from 1*s* to 4*p*. It has been shown that resonant diffraction at the *K* edge is valuable in determining charge and orbital ordering [20, 21], but due to the fact that the 4*p* states are probed, this information is rather indirect.

Figure 4 shows a rocking curve of the (016) and (106) reflections taken at ambient temperatures with the Pilatus I pixel detector at 7.7 keV, just below the Co *K* edge. The figure shows a sharp and a broad peak at almost the same position in reciprocal space, corresponding to the (016) and (106) reflections, respectively. The width of the (016) reflection is determined by the instrumental resolution. This reflection is already present with the high temperature Pmmm crystal structure. The (106) reflection, on the other hand, is a superlattice reflection sensitive to the doubling of the *a* axis, and it arises below the metal-insulator transition $T_{MI}\approx$350K, as shown in Figure 5. These two reflections appear at almost the same position in reciprocal space because of the *a*,*b*-twinning and the small *h* and *k* components. The width of the (106) reflection yields a correlation length of approximately 500Å. This finite correlation length is likely caused by the slight deviation from the ideal 5.5 composition of the oxygen content, which is found for the air annealed sample to be approximately 5.45. The missing oxygen ions will not only lead to stacking faults but also to different electronic states of the Co ions (electron doping). In addition, some of the octahedra will be replaced by square pyramids.

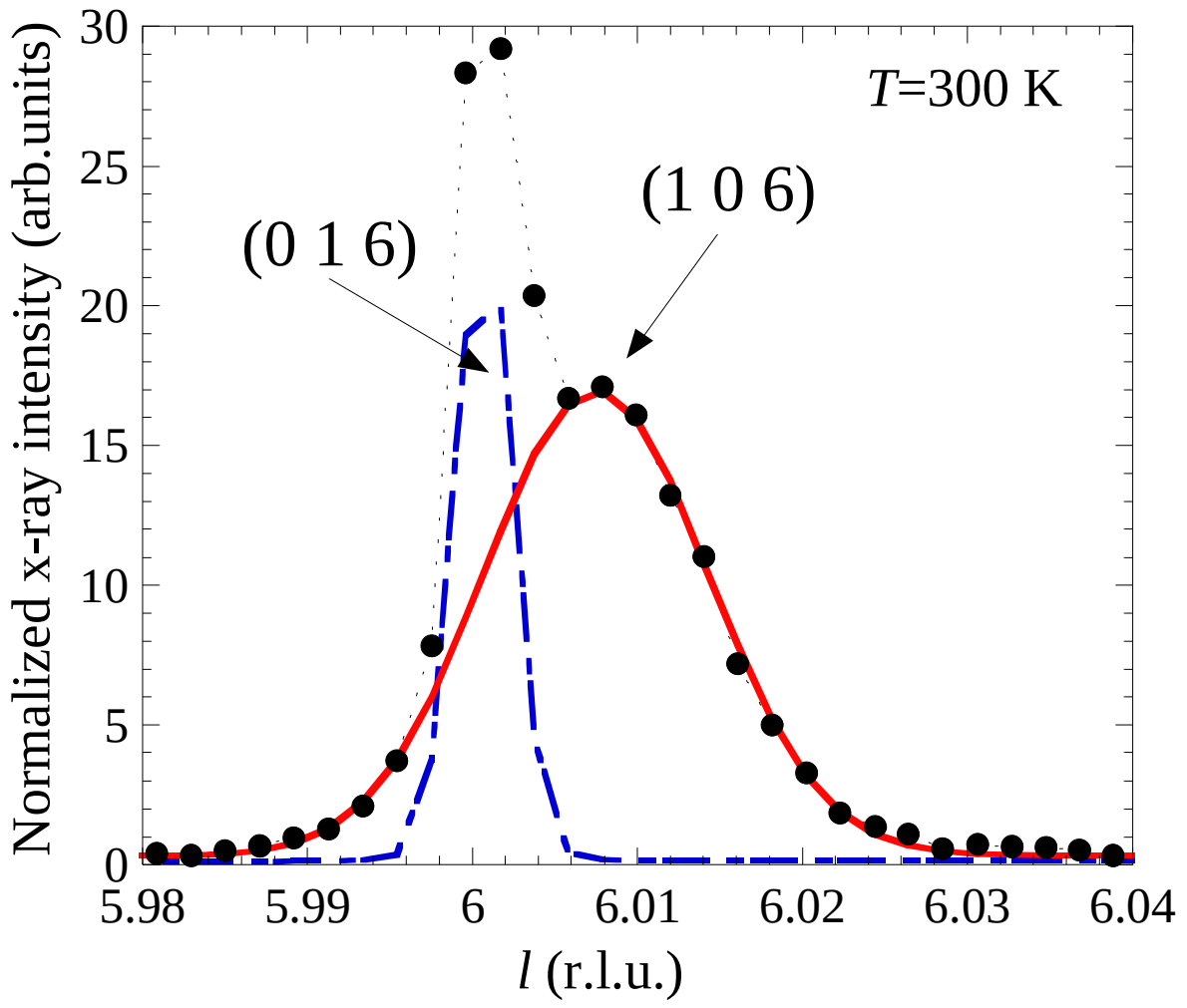

**Figure 4.** Rocking curve of the (106) and (016) reflections for twinned $GdBaCo_2O_{5.5}$, recorded with the two dimensional Pilatus I detector. The data were recorded below the Co *K* edge at 7.7 keV and at ambient temperature.

This assignment of the two reflections is supported by differences in their energy dependence across the Co *K* edge. The energy dependences of the (10*l*) and (01*l*) reflections recorded with incoming σ polarization and diffracted σ' polarization (σ–σ'), with *l* equal to 6 and 8, are shown in Figure 6. The (01*l*) reflections are sensitive to differences in scattering factor between the Co in the square pyramid and that of the Co in the octahedron. The (10*l*) reflections are sensitive to the electronic ordering among adjacent pyramids and octahedra.

The first group shows different energy profiles for *l*=6 and *l*=8: The (016) reflection shows an enhancement of intensity at 7.72 keV, whereas the (018) reflection shows a suppression. Moreover, the overall intensity of (018) is larger. This indicates that for the (018) reflection, the resonant scattering contribution is out of phase by π compared to the non-resonant scattering contribution, whereas for the (016) reflection it is in phase in the pre-edge region. In contrast, the superlattice reflections (106) and (108) have a similar energy dependency. The (108) intensity appears weaker than the (106), because this reflection was recorded further away from its maximum to avoid a contribution from the (018) reflection. No signal in the rotated polarization channel ($\sigma - \pi$') was found at this particular azimuthal angle, where the *a* axis (*b* of the twin) is approximately in the scattering plane.

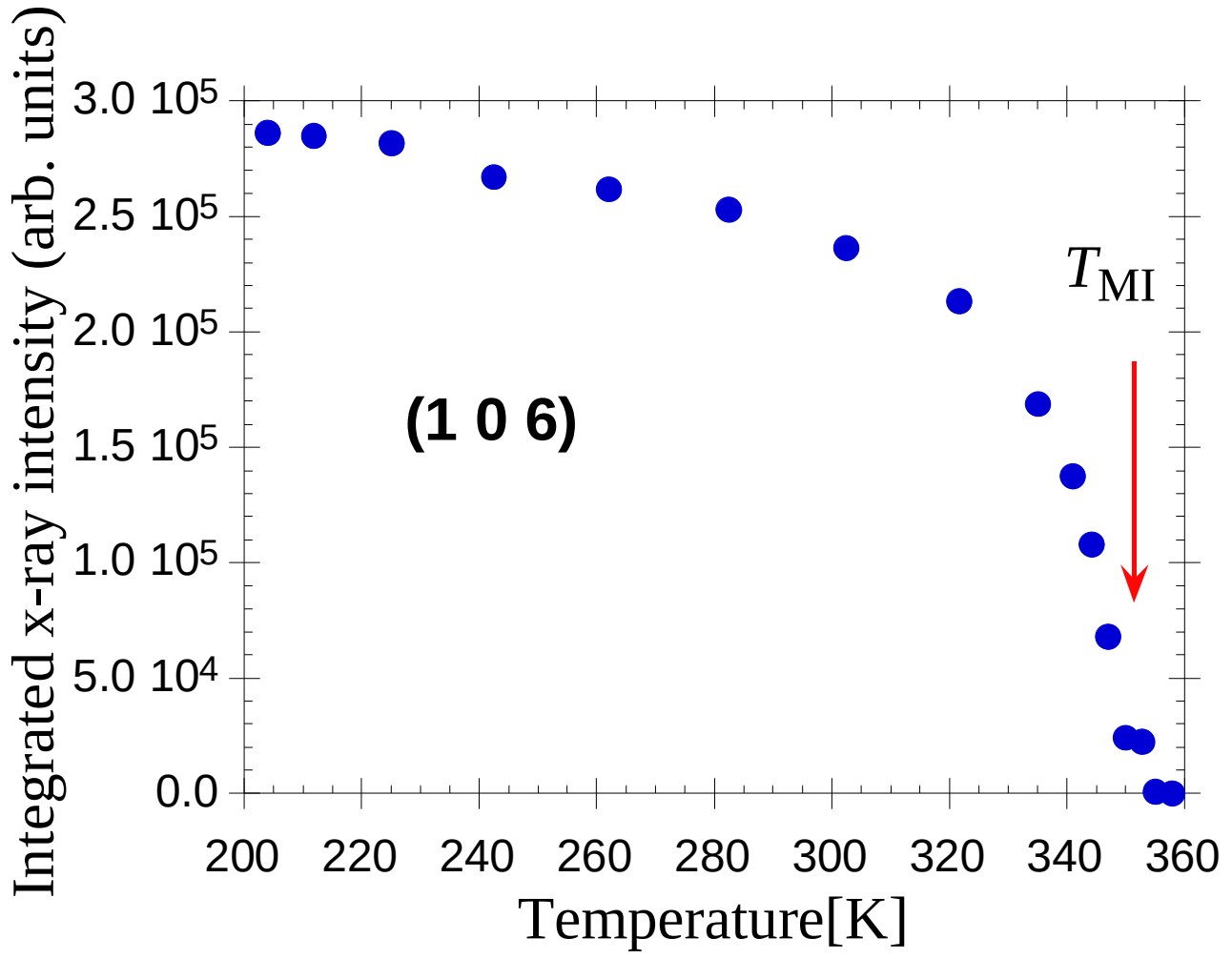

**Figure 5.** Temperature dependence of the (106) reflection, taken at 7.722 keV (Co *K* edge)

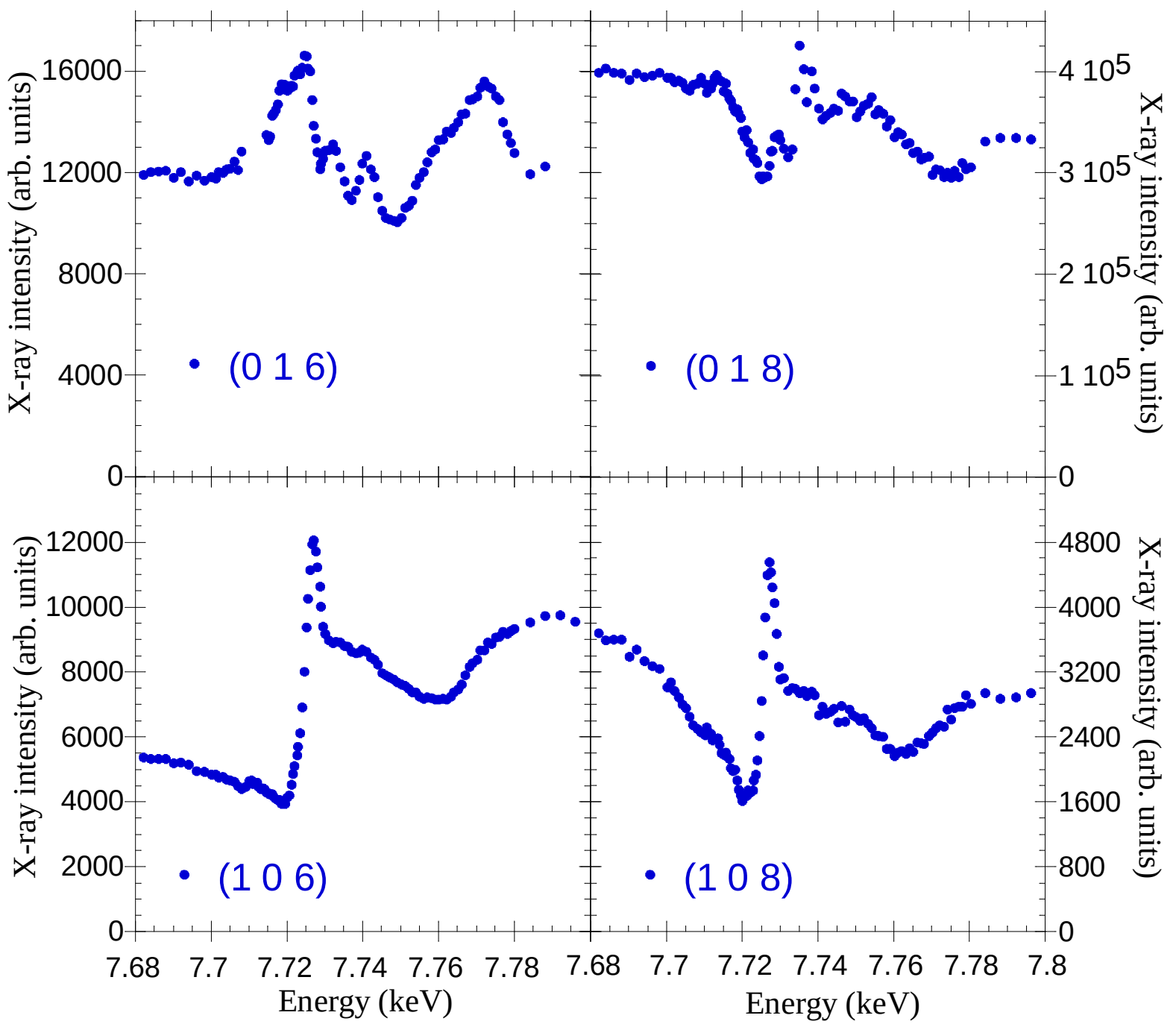

**Figure 6.** Energy dependence of (01*l*) and (10*l*) reflections, in the vicinity of the Co *K* edge, measured at ambient temperature with σ-σ’ polarization.

There are four sites occupied by the Co ions in the Pmma unit cell. The two pyramidal sites are labeled $Co_{Py1}$ and $Co_{Py2}$, and the two octahedral sites $Co_{Oc1}$ and $Co_{Oc2}$. Each site may have a different spin, charge and orbital state. The resonant part of the structure factor for the charge scattering of the (*h*0*l*) and (0*kl*) reflections is given by

$$F_{Co}^{(h0l)} \propto \left\{ f_{Py1}^{res} \sin(2\pi l \cdot z_{Py1}) + f_{Py2}^{res} \sin(2\pi l \cdot z_{Py2}) + f_{Oc1}^{res} \sin(2\pi l \cdot z_{Oc1}) + f_{Oc2}^{res} \sin(2\pi l \cdot z_{Oc2}) \right\} \quad (5)$$

$$F_{Co}^{(0kl)} \propto \left\{ f_{Py1}^{res} \cos(2\pi l \cdot z_{Py1}) + f_{Py2}^{res} \cos(2\pi l \cdot z_{Py2}) - f_{Oc1}^{res} \cos(2\pi l \cdot z_{Oc1}) - f_{Oc2}^{res} \cos(2\pi l \cdot z_{Oc2}) \right\} \quad (6)$$

Because the position z= $z_1 \approx 1\text{-}z_2$ for both the Co pyramidal and octahedral ions, Equation 5 can be approximated by $F_{Co}^{(h0l)} \propto \left( f_{Py1}^{res} - f_{Py2}^{res} + f_{Oc1}^{res} - f_{Oc2}^{res} \right) \sin(2\pi l \cdot z)$ and Equation 6 by $F_{Co}^{(0kl)} \propto \left( f_{Py1}^{res} + f_{Py2}^{res} - f_{Oc1}^{res} - f_{Oc2}^{res} \right) \cos(2\pi l \cdot z)$. Since *z* is close to ¼ for ions with both pyramidal and octahedral coordination, and if *l* is even, the resonant charge scattering structure factor is negligible for the (*h*0*l*) reflections, whereas it contributes to the (0*kl*) reflections. Moreover, the resonant charge structure factor for (0*kl*) changes sign between *l*=4*n* and *l*=4*n*+2, which is in agreement with the observed negative and positive interference with the non-resonant scattering, as discussed earlier. Because the resonant charge scattering for the (*h*0*l*) reflections is negligible, the resonant enhancement of these reflections must be caused by charge anisotropy between the Co ions. To analyze the anisotropies, we note that all Co sites have positions of the type I (1/4 0 *z*) or (1/4 1/2 z) or II (3/4 0 –*z*) or (3/4 1/2 -z). Positions I and II are related by a rotation by π around y or x, which therefore leads to a tensorial scattering component *T*, for which the structure factor has the form $\sum_{Co} 2i\left[T(I) - T(II)\right]$. Therefore, the resonant diffraction of the (*h*0*l*) reflections probes the difference in anisotropy of the Co ions at the four sites. The superlattice reflections recorded in this work are sensitive only to the anisotropy of the Co states along the *c*-axis.

In summary, the resonant hard x-ray diffraction experiments on the superlattice reflections occurring at $T_{MI}$ are sensitive to the orbital ordering and show a finite correlation length. However, first principle calculations, including a detailed tensorial analysis of the azimuthal angle dependences of additional superlattice reflections, are required to precisely determine the orbital ordering, and will be the topic for another study.

## V. RESONANT MAGNETIC SOFT X-RAY DIFFRACTION.

The magnetic (00½) reflection was recorded at the Co $L_3$ edge (777.5 eV), with polarization analysis of the scattered radiation. The scans corresponding to the four different polarization channels are shown in Figures 7a and 7b**.** The σ' and π' diffracted radiation was collected in two separate experiments, and the absolute intensities of these two data sets are different. The absence of Bragg diffraction in the σ-σ' channel provides direct evidence for the magnetic origin of the scattering, because resonant magnetic scattering is not allowed in this channel.

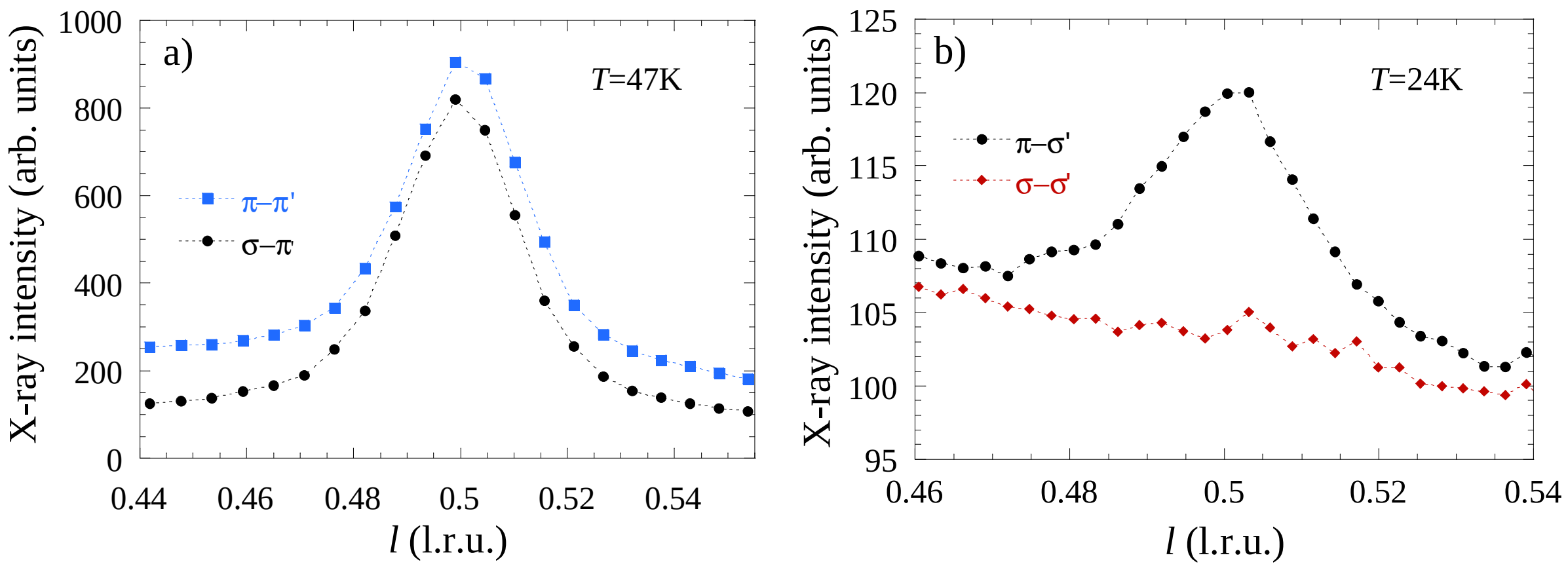

**Figure 7.** The magnetic (0 0 ½) reflection, measured at the Co $L_3$ edge, 777.5 eV, for the four different polarization channels; a) shows the signal collected in the π' detection channel, and b) shows the signal for the σ' detection channel. The absolute intensities for a) and b) are different.

## V.1 – COMPONENTS OF THE CO MAGNETIC MOMENTS.

The azimuthal angle (ψ) dependence was collected by rotating the sample around the (00½) reflection at the Co $L_3$ edge (777.5 eV) at T = 19 K. The ratio of the integrated diffracted intensity between π and σ radiation is shown in Figure 8 as function of the azimuthal angle ψ. Experimental artifacts due to misalignment, sample shape, etc, are cancelled out by evaluating $I_\pi/I_\sigma$, and such plots provide valuable information about the orientation of the magnetic moments [22]. The variation of the diffraction amplitude in an azimuthal scan at resonance is given by:

$$F(\theta)=\sum_{Kq}(-1)^q \mathrm{X}_{-q}^{K}(\theta)\sum_{q'} D_{q'q}^{K}(\gamma_0,\beta_0,\alpha_0)\Psi_{q'}^{K} \qquad (7)$$

The angles $\alpha_0$, $\beta_0$, $\gamma_0$ are related to the Euler angles of the rotation that aligns the Bragg vector $\tau$ along the $-a$ axis. $\Psi_q^K$ is the tensor associated with the electrons and is given by the expression $\Psi_q^K=\sum_d e^{iQ.d}\langle T_q^K\rangle_{E1,d}$ , in which the sum over $d$ runs over all the resonant ions in the unit cell, and where $d$ is the position of the ions within the unit cell. $D_{QQ'}^{K}(\alpha_0,\beta_0,\gamma_0)$ are the Wigner functions that correspond to the matrix elements of the rotations in the angular momentum representation [23], and $K$ is the rank of the tensor. A tensor with rank $K$=1 represents the origin of magnetic scattering.

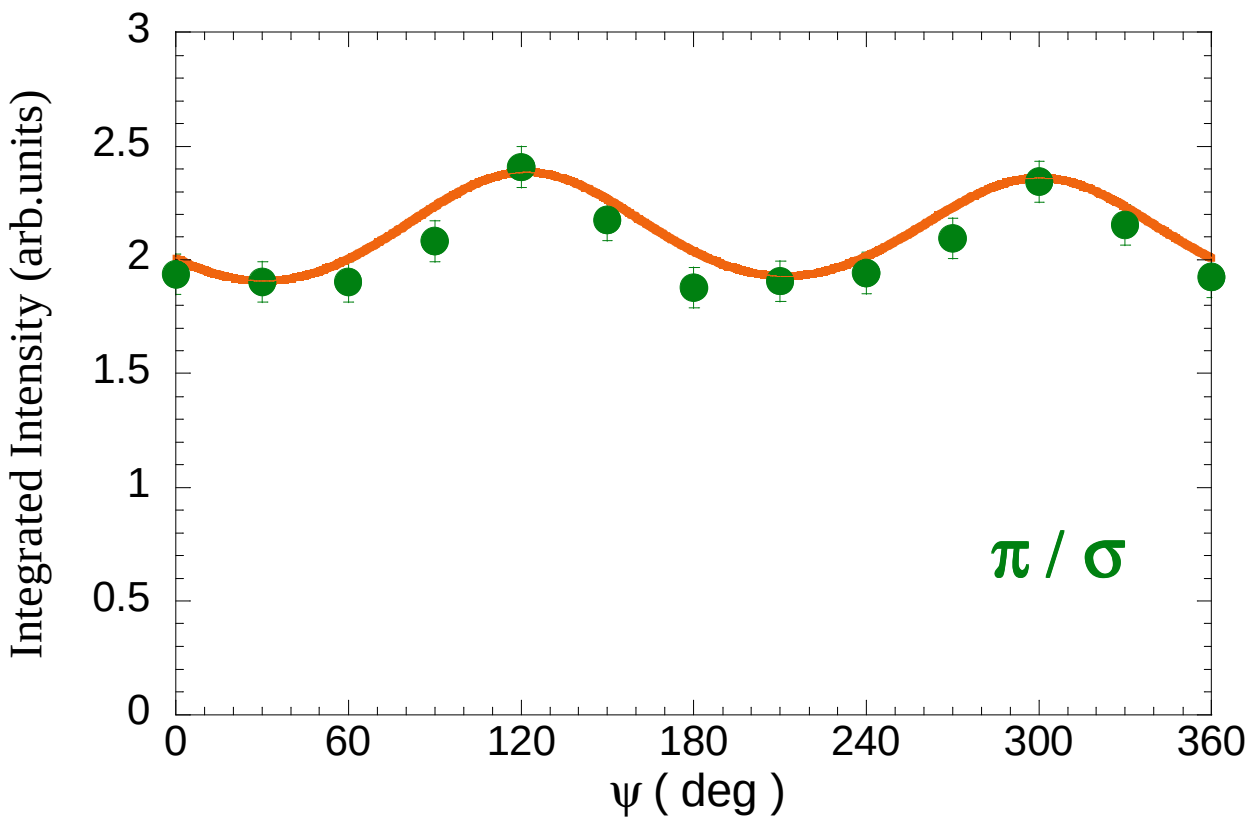

**Figure 8.** Ratio of the integrated x-ray intensity of the (00½) reflection taken with π and σ incident radiation at T= 19K, as function of azimuthal angle. The fit, as described in the text, is shown with a continuous curve.

Figure 8 illustrates that $I_\pi/I_\sigma$ >1 for all ψ, which indicates a strong π-π' contribution at all azimuthal angles, as confirmed with the polarization analysis (see Figure 7). In addition, a clear two fold periodicity is observed.

Because our single crystal is twinned, we have to consider individual weighting factors for the two crystallographic domains. The hard x-ray diffraction measurements have demonstrated equal populations of each domain type, within the probing volume of 200x200x40 $\mu m^3$. If the magnetic moments are in the *ab* plane, equal domain populations imply that $I_\pi/I_\sigma$ (ψ) is constant. Yet Fig. 8 illustrates a modulation, and hence indicates that the domain populations are not exactly equal. This is likely due to the limited penetration depth of the soft x-rays at resonance, which is of the order of 100 nm at the Co *L*-edge. Such an unequal domain population is not uncommon in resonant x-ray scattering experiments, as found, e.g., in $V_2O_3$ [24].

We consider two different domains, labeled **A** and **B**, where the *a*, *b* axes in domain **A** are perpendicular to the *a'*, *b'* axes in domain **B**. The tensor associated with the electronic order becomes in this case $\Psi^K_{\pm q} = \mathbf{A} \pm \mathbf{B}$. We obtain for the diffraction amplitude:

$$F(\theta) = \sum_q F_q(\theta) = \mathrm{X}^K_{-q}(\theta)(-1)^q \cdot (-1)^q \cdot \sum_p e^{ip\psi} d^K_{qp}\left(\frac{\pi}{2}\right) \cdot \left(A_p + B_p\right) \tag{8}$$

where the rank of the tensor *K*=1 (magnetic), $-k \le q \le k$, and *p* = 0, ±1. Expressing **A** and **B** in terms of the component of the magnetic moment along the *x* and *y* axis**:**

$$\Psi^1_0 \equiv m_z \tag{9}$$

$$\Psi^1_x = \frac{1}{\sqrt{2}}\left(\Psi^1_{-1} - \Psi^1_{+1}\right) \propto m_x \tag{10}$$

$$\Psi^1_y = \frac{i}{\sqrt{2}}\left(\Psi^1_{-1} + \Psi^1_{+1}\right) \propto m_y \tag{11}$$

we deduce for the intensities in the four different polarization channels for a single domain:

$$I_{\pi-\sigma'} = \left| \left( \frac{i}{\sqrt{2}} \right) \cdot \{ (m_x \cdot \sin\varphi + m_y \cdot \cos\varphi) \cdot \cos\theta + m_z \cdot \sin\theta \} \right|^2 \quad (12)$$

$$I_{\sigma-\pi'} = \left| -F_{\pi'\sigma}(-\theta) \right|^2 \quad (13)$$

$$I_{\pi-\pi'} = \left| \frac{i}{\sqrt{2}} \cdot \sin(2\theta) \cdot \{ -m_x \cdot \cos\varphi + m_y \cdot \sin\varphi \} \right|^2 \quad (14)$$

$$I_{\sigma-\sigma'} = 0 \quad (15)$$

The parameters $m_x$, $m_y$ and $m_z$ reflect the sum of the *x*, *y* and *z* components of the magnetic moment, weighted by the phase factor of the ion position in the structure factor, respectively. These intensities describe any magnetic structure with the given ordering wavevector (0 0 1/2) in $R\mathrm{BaCo_2O_{5.5-\delta}}$.

We fit $\frac{I_\pi}{I_\sigma} = \frac{(I_{\pi-\sigma'} + I_{\pi-\pi'})}{(I_{\sigma-\pi'} + I_{\sigma-\sigma'})}$ to the experimentally observed ratio and determine the magnitude of the magnetic moment along each axis and the population of the two domains. Because the presence of the two twin domains, we can determine only the relative magnitudes of $m_x$, $m_y$ and $m_z$. We set $m_x = 1$ and obtain $m_y = 1.66 \pm 0.02$, $m_z = 0.006 \pm 0.001$ and a relative domain population $\alpha$ = **A**/**B** = 0.550 ± 0.001. The same result is found for $\alpha = 0.45$ and $m_y = -1.66 \pm 0.02$. The result of this fit is illustrated in Figure 8. Assuming a collinear magnetic structure, both solutions correspond to magnetic moments aligned in the *ab* plane and rotated 30º ± 5º with respect to the *a* or *b* axis, respectively. However, non-collinear models are not excluded. Magnetic moments aligned along the *a* or *b* axis are, however, inconsistent with our observations.

The rotation of the magnetic moment in the *ab* plane is similar to that proposed for $\mathrm{TbBaCo_2O_{5.5}}$ resulting from NMR measurements [11]. The Co moments in the $\mathrm{CoO_5}$ pyramids lie in the *ab*-plane and are canted at an angle ~ 45º from the *a*-axis at low temperatures. This study reported a variation of the orientation of the magnetic moments as a function of temperature. A tilting of the moments away from the *a* axis contradicts the magnetic structures proposed for $GdBaCo_2O_{5.5}$ [12], $NdBaCo_2O_{5.47}$ [9] and $TbBaCo_2O_{5.5}$ [6]. In all three cases, the magnetic moment was proposed to be aligned along the *a* axis. This proposal may have been triggered by the fact that for a monodomain single crystal, the moment in the ferrimagnetic phase points along the *a* axis [12]. Given a magnetic moment along the *a* axis, it has been proposed that the AFM to FM transition is due to a change from an FM to an AFM exchange interaction and reflects a simple spin flip. Moreover, simple symmetry arguments indicate that the easy axis in an orthorhombic system is along a single crystallographic direction. In a complicated material such as this layered cobaltite, however, competing exchange interactions, and consequently more complicated magnetic structures, e.g. non collinear with large canting angle, are viable.

In summary, the azimuthal angle dependence of the magnetic reflection gives clear evidence, that the magnetic moments are not lying parallel a single crystallographic axis, either *a* or *b*.

## V.2 – ELECTRONIC STATES OF THE MAGNETIC CO IONS.

The energy dependence of the (00½) reflection, recorded in the vicinity the Co $L_{2,3}$ edges at 99 K and with π' diffracted radiation, is shown in Figure 9. The spectra are characterized by three features at the $L_3$ edge (A, B, C) and two features at the $L_2$ edge (D, E). For the π-π' intensity, features indicated with (*) originate from the strong anomalous reflectivity at the Ba $M_{4,5}$ edges. These features are confirmed to be non-magnetic by θ/2θ scans, for which the reflectivity is only smoothly varying.

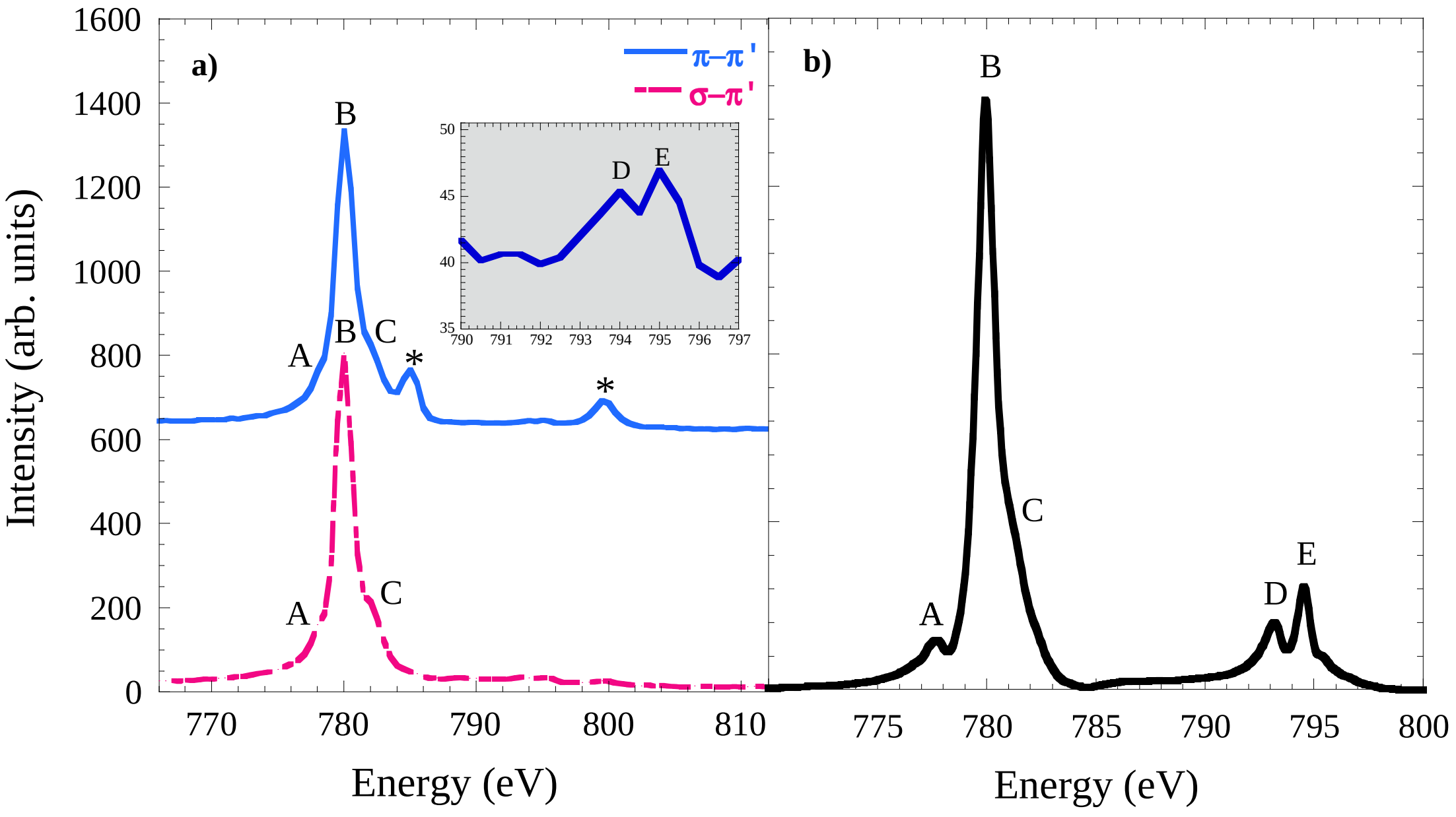

**Figure 9.** a) Energy dependence of the magnetic (0 0 1/2) reflection in the vicinity of the Co $L_{2,3}$ edges, recorded at 99 K with polarization analysis (left graph) compared with b) theory.

The computer code Hilbert ++ [25] has been used to calculate the energy dependence of the magnetic (0 0 1/2) reflection. The $Co_{Oc}$ and $Co_{Py}$ sites were modeled independently with $CoO_5$ and $CoO_6$ clusters and the structural information from the Pmmm structure of Ref. [7]. The Hilbert space is spanned by the states having 6+*n* electrons in the 3*d* shell and *n* holes in the oxygen 2*p* orbitals. The excited system is described by the states having one hole in the Co 2*p* shell, 7+*n* electrons in the 3*d* shell and *n* holes in the oxygen orbitals. The atomic parameters are shown in Table 2.

| | $F_{dd}^2$ | $F_{dd}^4$ | $\zeta_{3d}$ | $F_{pd}^2$ | $G_{pd}^1$ | $G_{pd}^3$ | $\zeta_{2p}$ | $F_{dd}^0$ | $F_{pd}^0$ |
|---|---|---|---|---|---|---|---|---|---|
| Ground state | 10.13 | 6.34 | 0.074 | 5.7 | 4.2 | 2.4 | 9.39 | 4.0 | 4.4 |
| Excited state | 10.17 | 6.36 | 0.075 | 5.7 | 4.2 | 2.4 | 9.4 | 4.0 | 4.4 |

**Table 2.** Atomic parameters for the theoretical calculation. *F* stands for a direct Slater integral, *G* for a exchange Slater integral and ζ for a spin-orbit coupling.

The Slater integrals have been calculated using the Cowan code [26] and have been rescaled by a factor of 0.8 to take into account screening effects. The Slater monopole integrals $F_{dd}^0$ and $F_{pd}^0$, which are strongly screened, have been fixed in agreement with Ref. [27]. The Slater-Koster σ and π parameters for 3*d*-2*p* hybridization have been fixed to 2.3 and 0.9 respectively. The hybridization strength has been rescaled, proportional to $\left(\frac{2}{r}\right)^3$, where *r* is the bond length. The crystal field (CF) has been set to zero. The oxygen orbital energy has been chosen such that without hybridization the oxygen 2*p* electrons are 4.5 eV below the first free 3*d* level.

This computation shows that the Co spin state in the $CoO_5$ cluster is HS and that in the $CoO_6$ cluster is LS. Moreover, the single crystalline anisotropy defines the easy axis of the $CoO_5$ to be the *a*-axis and that of the $CoO_6$ to be the *b*-axis. Upon reducing the hybridization between Co and O, HS is found for the $CoO_6$ cluster, while adding a small CF stabilizes LS. The calculated X-ray absorption shows the difference between LS and HS (Figure 11), but is insensitive to the CF.

The calculations for the magnetic reflection, normalized and corrected by the calculated absorption, are also shown in Figure 9. Good agreement with experiment is observed: the calculated features (A-C) at the Co $L_3$ edge and (D, E) at the $L_2$ edge have approximately the observed energy splitting (see inset Figure 9). However, the intensity ratio between the Co $L_3$ and $L_2$ edge is significantly larger in the experiment compared to the calculations, indicating an underestimation of the orbital magnetic moment. This is further discussed in Section VI.

In summary, the energy dependence of the magnetic ( 0 0 ½ ) reflection is qualitatively described by HS $Co^{3+}$ in the square pyramid.

## V.3 – TEMPERATURE EVOLUTION OF MAGNETIC PHASES

The temperature dependence of the magnetic (0 0 1/2) reflection is shown in Fig. 10, together with the magnetization recorded in an applied magnetic field of 0.01 T parallel to the *ab* plane. Four different regions can be distinguished. Below $T_1$ = 275 K, region $III_A$ is characterized by an increase in magnetization with decreasing temperature. Below T=245 K, region $III_B$ begins where the magnetization decreases. Region $III_C$ is characterized by the appearance of the magnetic (0 0 ½) reflection at $T_2$=230 K. Region IV begins when the magnetization decreases more strongly and the intensity of the (0 0 ½) reflection increases more strongly. Here the roman letters refer to the generic phase diagram of Figure 1.

Figure 10 can be interpreted as follows: At $T_1$ = 275 K, the system undergoes a magnetic transition, evidenced by the strong increase in magnetization. This transition was previously reported to coincide with the appearance (note we use the $2a_px2a_px2a_p$ cell) of the (1 1 1) antiferromagnetic (AFM) reflection in neutron powder diffraction experiments in $NdBaCo_2O_{5.47}$ and $TbBaCo_2O_{5.5}$ [6, 7]. This suggests that the spins are aligned antiferromagnetically and canted in region $III_A$ with little or no temperature dependent change in the canting angle. It is the canting of the spins leading to the ferromagnetic component observed in the magnetization. This is also consistent with the results obtained from cluster calculations (section V.2) and with the soft x-ray azimuthal angle dependence, which also indicates a spin canting in phase IV (section V.1).

For T≤245 K, the susceptibility turns around and starts to decrease. The appearance of a new magnetic phase that coexists with phase $III_A$ is unlikely, because no evidence of such a mixed phase has been reported so far at this temperature. The decrease of the FM component in region $II_B$ can be explained by a change in the canting angle of the magnetic moments, i.e. the magnetic moments rotate clockwise and anticlockwise, reducing the ferromagnetic component in spite of a further increase of the unique magnetic moment.

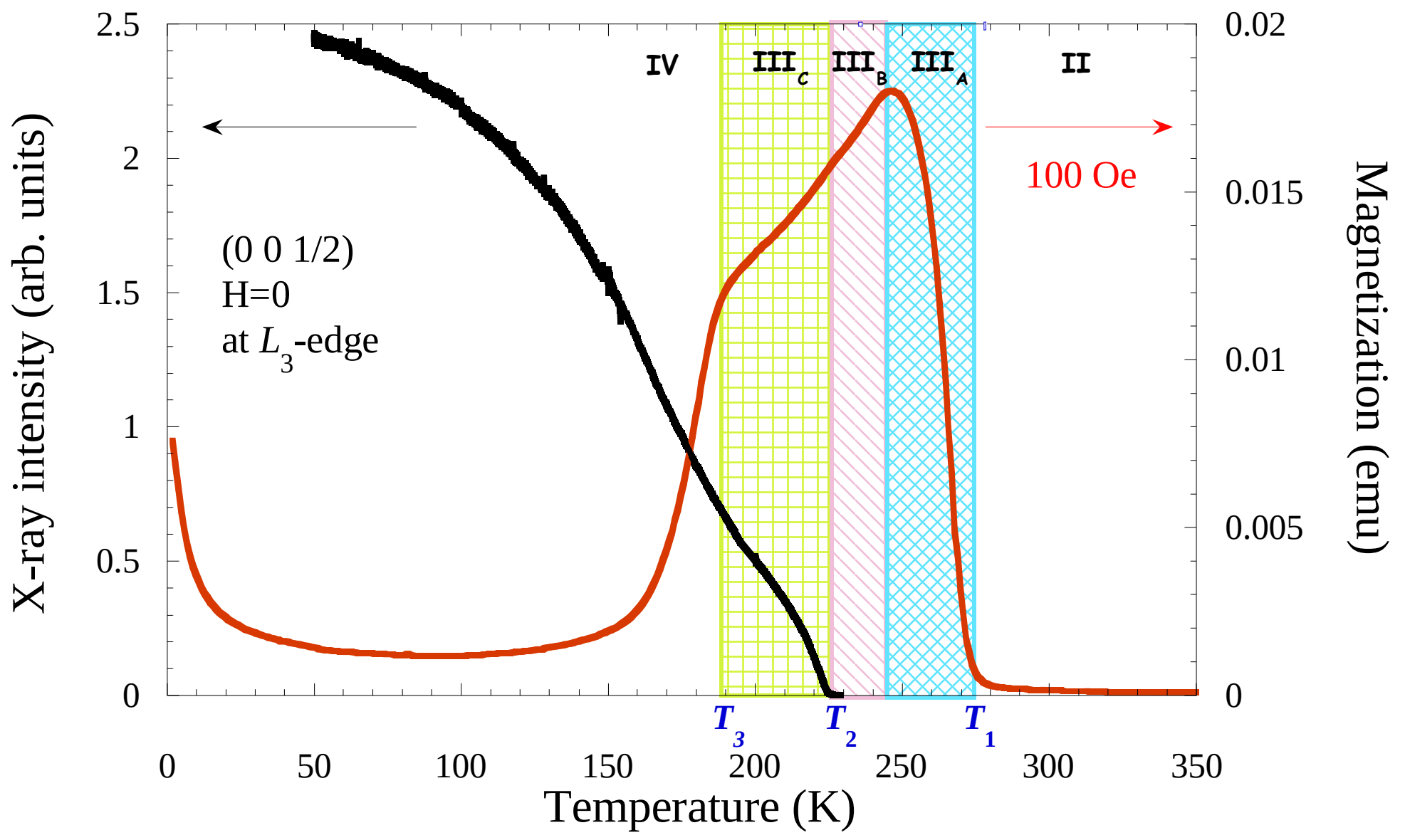

**Figure 10.** Temperature dependence of the AFM (0 0 1/2) reflection measured at the Co $L_3$ edge (777.5 eV) and magnetization of the single crystal in an external field of 100 Oe applied in the x, y plane.

Below $T_2$=230 K region **III**$_C$, pure antiferromagnetic (AF) phase appears, indicated by the observation of the AF (0 0 ½) reflection. The FM component does not disappear immediately, and the temperature dependence of the AFM reflection (0 0 ½) does not show a regular order parameter behavior in the region of 210 – 150 K, but rather a slight convex temperature dependence. This indicates a coexistence of two magnetic phases, due to the first order nature of the transition. Below T = 190 K (phase **IV**), the FM contribution strongly decreases, and the slope of intensity of the (0 0 1/2) AF reflection strongly increases. This is consistent with the neutron diffraction results on $NdBaCo_2O_{5.5}$ [6] and $TbBaCo_2O_{5.5}$ [7], where the intensity of the (1 1 1) reflection decreases in this region.

In summary, the temperature dependence of the susceptibility and the magnetic ( 0 0 ½ ) reflection suggests that the magnetic moments are canted in all phases and that the canting angle is temperature dependent, at least in phase III.

## VI. SPIN-STATE AND ORBITAL MAGNETIC MOMENT.

X-ray absorption data recorded at the Co $L_{2,3}$ edges of $GdBaCo_2O_{5.5-x}$ are shown in Figure 11. The absorption around 780 eV and 795 eV corresponds to the Co $L_3$ and $L_2$ edges, respectively. The sharp and strong absorptions at 785 eV and 800 eV correspond to the $M_5$ and $M_4$ edges of Ba. The Ba $M_{4,5}$ edges probe the empty 4*f* shell of the $Ba^{2+}$ ions and are irrelevant for the electronic properties of the material. This is proven by the absence of resonant (0 0 ½) diffraction at the Ba $M_{4,5}$ edges. The Co $L_3$ edge has three shoulders, indicated by A, B, C, and a small pre-edge feature labeled *. The latter is due to a small amount of $Co^{2+}$ in the sample, resulting either from a small oxygen deficiency or from a surface reduction effect. The $L_2$ edge also exhibits three features, indicated by D, E, F, although the shoulder on the low energy side is less pronounced.

The recorded XAS spectrum is compared with the theoretical calculations (section V.2) in Figure 11. There is reasonable agreement with the computational result, which assumes that $Co^{3+}$ exhibits a LS for the octahedral environment and a HS for the square pyramidal environment [28].

In Figure 11, the calculation for half filling of the $CoO_6$ octahedra with HS state is shown. The intensity of feature C is significantly reduced compared to the calculation where all $CoO_6$ octahedra exhibit a LS state. Therefore the observed intensity of feature C excludes spin state ordering in the *ac* plane of the octahedron. However, a small amount of $Co_{Oc}$ could be HS state, due to missing oxygen.

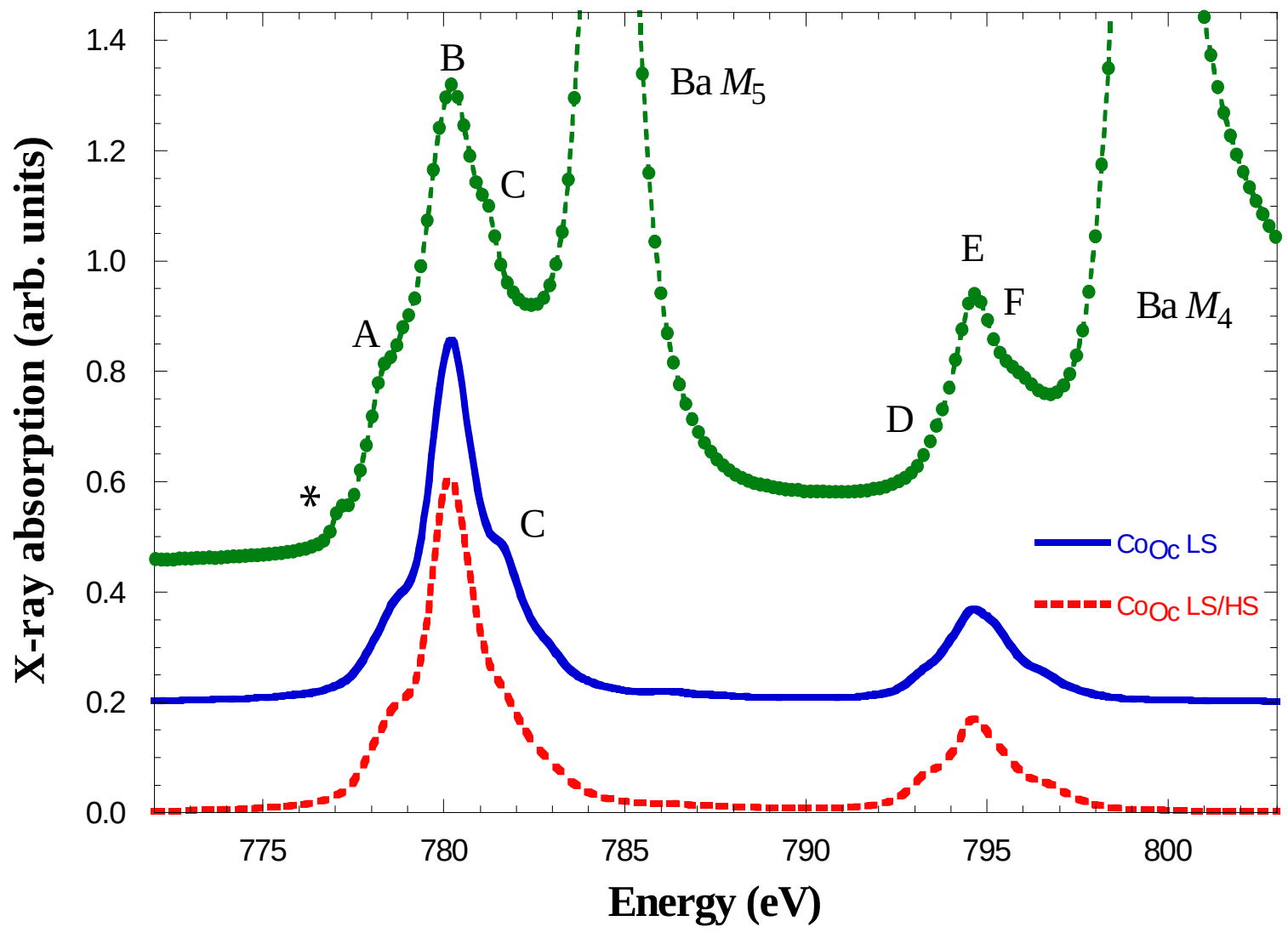

**Figure 11.** X-ray absorption spectra of $GdBaCo_2O_{5.5-x}$ in the vicinity of the Co $L_{2,3}$ edges, taken at 200K (top). The lower two curves show the theoretical calculations of the absorption edge with pyramidal Co in the HS state and octahedral Co in the LS state (solid curve) and with half of the octahedral Co in the HS state and the other half in LS (dotted curve). A to F indicate various specific features in the spectra, and * is presumably a contribution from $Co^{+2}$.

Figure 12 shows x-ray magnetic circular dichroism data collected at the Co *L* edges. The observed XMCD signal is rather weak and in the order of 1%. The $L_3$ edge XMCD signal is much larger than that observed at the $L_2$ edge, which is consistent with the resonant diffraction data (Figure 9). No XMCD signal is detected at the Ba $M_{4,5}$ edges (*d*-*f* transition), in accordance with a fully unoccupied character of the Ba 4*f* states. Resonant diffraction experiments and XMCD probe the same magnetic scattering factor. Therefore the square root of the integrated intensity in resonant scattering is compared with the integrated XMCD intensity. The ratio between the $L_3$ and $L_2$ signals is equal for both techniques, under the assumption a) that the XMCD signal does not change sign at each edge and b) that the Co electronic states for the ferromagnetic and antiferromagnetic components are the same. Whereas assumption b) is likely valid, a) is usually not valid, as it is common that XMCD signals of transition metal oxides change in sign at the 3*d*-ion $L_3$ edge [28].

However, Figure 12 shows no significant positive XMCD signal at the $L_3$-edge, which therefore allows such comparison.

The integrated intensity ratio between the Co $L_3$ and $L_2$-edge XMCD signal equals 8.1 and is slightly smaller than the ratio of the square roots of the resonant diffracted intensities, which equals 8±1 and 10±1.5 for $\pi$ and $\sigma$ incident radiation, respectively. It is also interesting to apply the sum rules to extract the ratio of orbital and spin magnetic moments. For the $Co^{3+}$ HS in the square pyramid, the magnetic dipole term in the sum rule is not necessarily negligible. However, the total integrated XMCD intensity is relatively large and indicative of a significant orbital magnetic moment. Assuming a negligible magnetic dipole term, the ratio between the orbital and the spin magnetic moment is approximately $L_z/S_z \approx 0.6$. This ratio is very similar to $L_z/S_z \approx 0.5$ found for $LaCoO_3$ [29], where the large orbital magnetic moment is attributed to the HS state of the $Co^{3+}$ ions. This finding clearly indicates that, to be meaningful, an extraction of the spin state from effective magnetic moments obtained from magnetization or neutron diffraction experiments must take into account the large contribution from the orbital magnetic moment. A roughly 50/50 mixture of LS/HS in the sample has significant impact on the magnetic properties, as it would directly indicate that there is no magnetic moment on the octahedrally coordinated LS $Co^{3+}$. This is in good agreement with the results of a soft x-ray absorption study on the $CoO_5$ pyramidal-coordinated $Sr_2CoO_3Cl$, where the $Co^{3+}$ was also found to be in the HS [28]. This LS/HS ratio between the two sites is also supported by the NMR results of Ref. [9]. However, these findings are in contrast to an IS for the $Co^{3+}$ with square pyramidal coordination in the $R$BaCo$_2$O$_{5.5}$ system, which is, to our knowledge, assumed by all other studies.

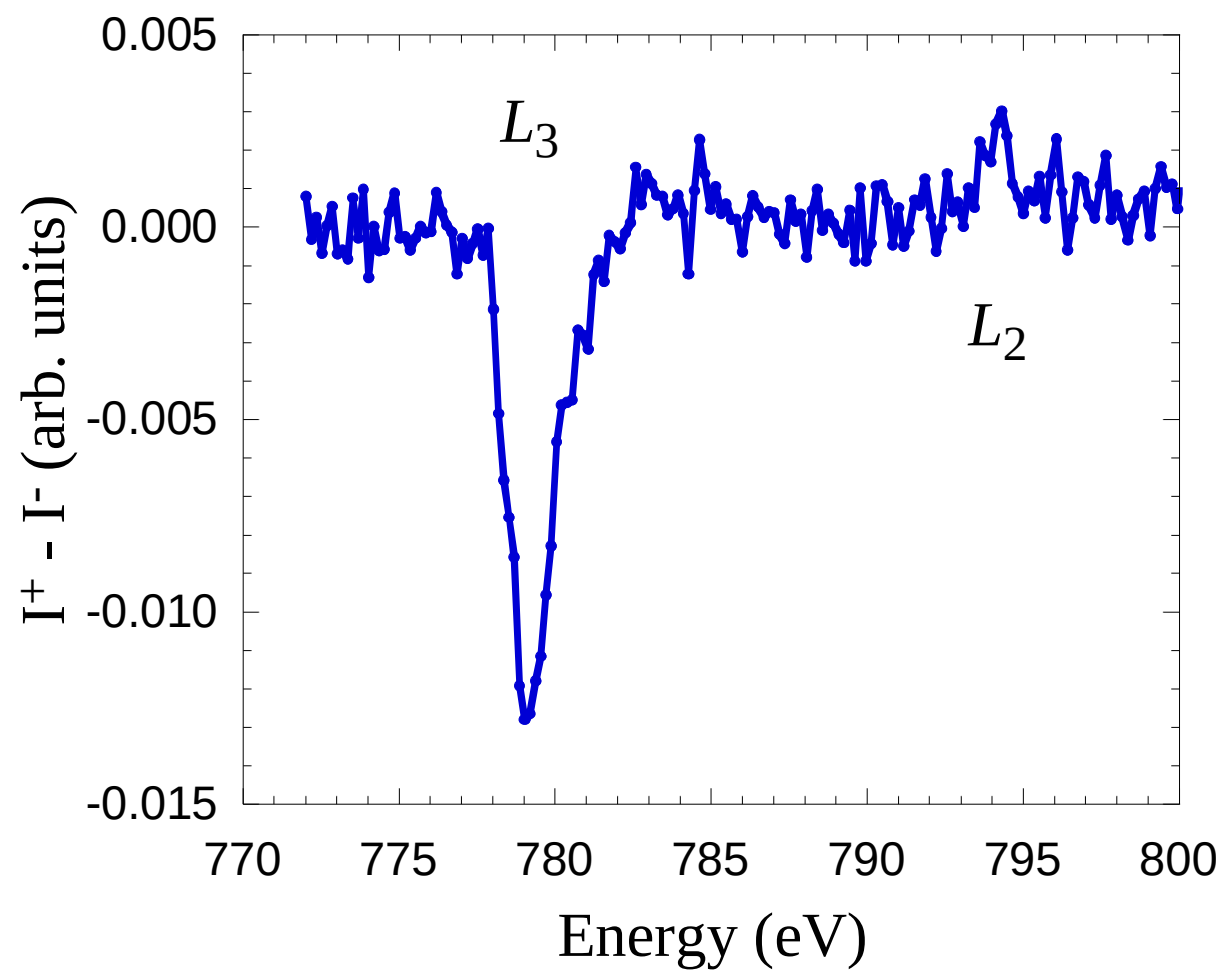

**Figure 12.** X-ray magnetic circular dichroism (XMCD) spectra of $GdBaCo_2O_{5.5-\delta}$ at the Co $L_{2,3}$ edges taken at 200K and in a field of 1.9T.

The XMCD signal at the Co $L_3$ edge is shown in Figure 13 as a function of applied magnetic field. A clear hysteresis effect is observed. An XMCD signal is also found at the Gd $M_5$ edge at 1183.2 eV, but it does not exhibit hysteresis (not shown). This indicates that the Gd magnetic moment is induced by the applied magnetic field and is due to the paramagnetic response of the Gd sublattice. This confirms that the ferromagnetic component of the magnetization arises solely from the Co magnetic moments.

In summary, the x-ray absorption results are best described by a 50% / 50%, LS/HS spin state distribution suggesting a LS state of the Co in the octahedra. The XMCD data indicates that the Co magnetic moments have a significant orbital moment contribution, important for the extraction of spin states from magnetization and neutron diffraction data.

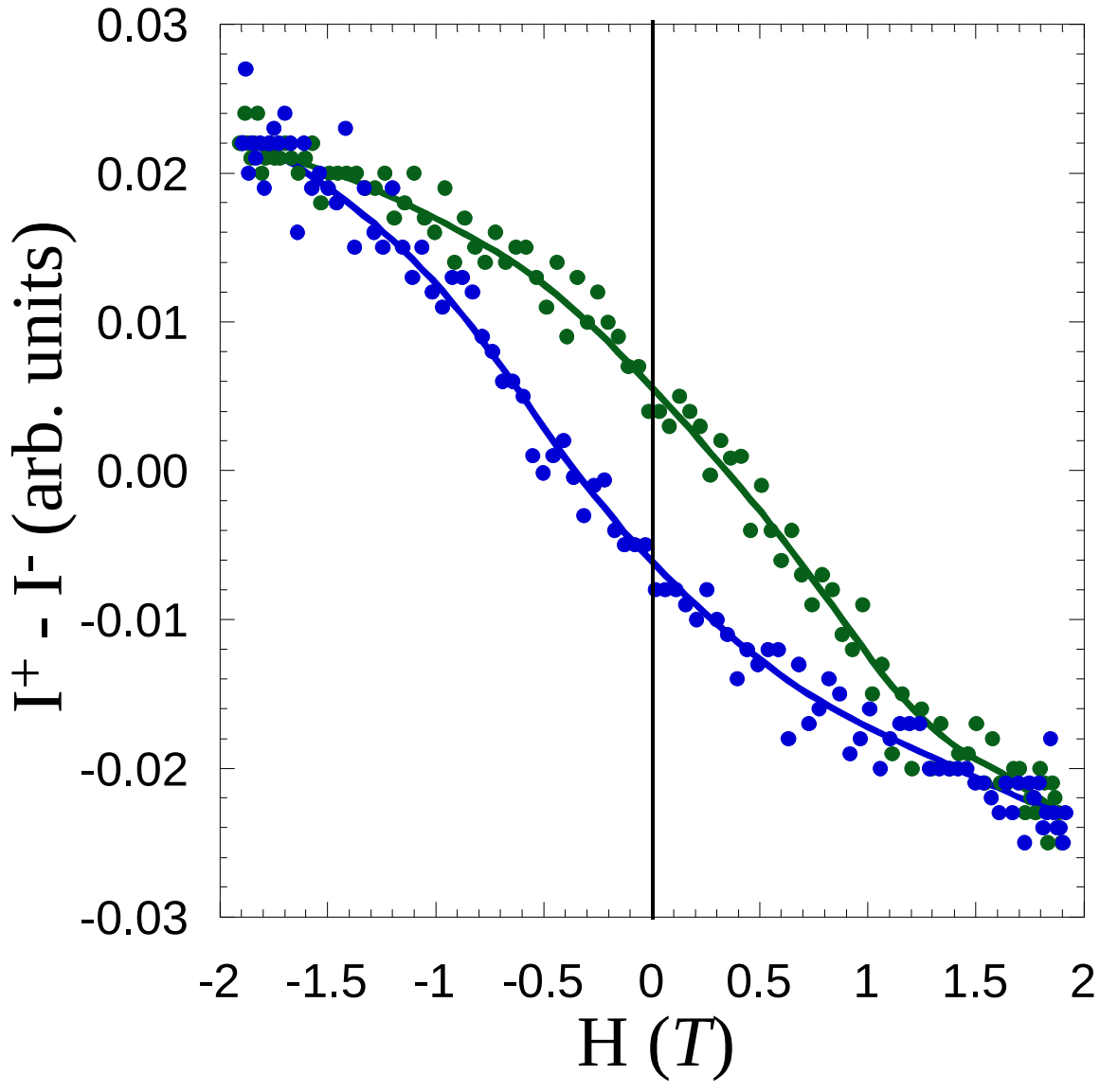

**Figure 13.** Hysteresis of the XMCD intensity, recorded at the Co $L_3$ edge at 779 eV at 200K. The curves are guides to the eye.

## VII. DISCUSSION AND CONCLUSIONS

We have performed non-resonant x-ray diffraction, resonant soft and hard x-ray magnetic diffraction, soft x-ray absorption spectroscopy and XMCD measurements to clarify the electronic and magnetic states of the $Co^{3+}$ ions in $GdBaCo_2O_{5.5}$. The XAS data are well described by a model where half of the $Co^{3+}$ ions are in the HS state and the other half are in the LS state. This is consistent with cluster calculations that find a $Co^{3+}$ HS state in the pyramidal sites and a $Co^{3+}$ LS state for the octahedral sites. The structural distortion shows alternating elongations and contractions of the pyramids and indicates that the metal insulator transition is associated with orbital order of the $Co_{Py}^{3+}$ HS state, with four electrons in the $t_{2g}$ orbitals. Due to the degeneracy of the $t_{2g}$ states, the fourth electron can either occupy an *xy*, *xz* or *yz* state. The observed elongations therefore indicate an alternating ordering of *xz* and *yz* orbitals along the *a* and *c* axes. A site distortion is found for the $Co^{3+}$ ions at the octahedral sites, but for the LS state of $Co_{Oc}$, no orbital order is possible, as $t_{2g}$ states are filled and the $e_g$ states are empty. Therefore, the size distortion is just the response of the orbital order in the pyramidal sites. Note that a charge order would strongly affect the L-edge XAS, and can therefore be excluded. Moreover, we note here that in case the pyramidal Co would be IS, it would reflect either a $x^2$-$z^2$ / $y^2$-$z^2$ or $3r^2$-$x^2$ / $3r^2$-$y^2$ orbital order. These interpretations are simplifications, because the hybridization of Co 3*d* and oxygen 2*p* states, directly visible in the ordering of the oxygen 2*p* states in manganites [30], is likely to be present.

The orbital ordering has significant impact on the magnetic ordering. The magnetic moments prefer to order perpendicular to the *xz* and *yz* orbitals and consequently will not align along a single crystalline axis, but rather alternate between the *a* and *b* axes. The azimuthal scan of the magnetic (0 0 1/2) reflection indicates that the sum of the magnetic moment components along the *b* axis is 1.66(2) times larger than that along the *a* axis. For a collinear magnetic model, this implies that the magnetic moments align in the *ab* plane, 30 degrees away from the *a* or *b* axis. However, the pure LS state character of the octahedral Co, with orbital order and a doubling of the magnetic unit cell along the *c* axis, rather supports a non-collinear magnetic structure, with moments aligned approximately perpendicular to one another, along the *c* axes. Both unequal magnetic moments and alignments tilted away from the crystallographic axes are possible. If the magnetic moments are all of equal magnitude, this will imply that the moments are tilted away from the *a* and *b* axes, on the average by 30 degrees. Alternatively, if the moments are aligned along the *a* and *b* axes then there must be two (or more) magnetic moments in the material, one 1.66 times larger than the other which is unlikely.

In conclusion, our experimental results indicate a HS state for the $Co^{3+}$ in the pyramids, with orbital order in the $t_{2g}$ states in the *ac* plane. Further, the $Co^{3+}$ is in a LS state in the octahedral site. The magnetic structure is proposed to be non-collinear, possibly with different magnetic moments and/or with moments aligned in the *ab* plane, tilted from the *a* and *b* axes. In addition the magnetic moments show a large orbital component.

## Acknowledgements

We wish to thank J.M. Tonnerre for supporting the establishment of the RESOXS endstation, S. Lovesey for his considerable help during the azimuthal scans analysis and Arantxa Fraile-Rodriguez, for help analyzing the XMCD data. Financial support from the Swiss National Science Foundation is gratefully acknowledged. The experiments were in part performed at the SLS of the Paul Scherrer Insitut, Villigen, Switzerland, and we wish to thank the beamline staff of X11MA for their excellent support.